# Prototype of time digitizing system for BESIII endcap TOF upgrade*

CAO Ping(曹平)[1,2; 1)] SUN Wei-Jia(孙维佳)[1,2] JI Xiao-Lu(季筱璐)[1,3;2)]
FAN Huan-Huan(范欢欢)[1,2] WANG Si-Yu(王思宇)[1,2] LIU Shu-Bin(刘树彬)[1,2] AN Qi(安琪)[1,2]

[1] State Key Laboratory of Particle Detection and Electronics
[2] Anhui Key Laboratory of Physical Electronics, Department of Modern Physics, University of Science and Technology of China, HeFei 230026, China
[3] Institute of High Energy Physics, Chinese Academic of Science, Beijing 100049, China

**Abstract:** The prototype of time digitizing system for the upgrade of BESIII endcap TOF (ETOF) is introduced in this paper. The ETOF readout electronics has a formation of distributed architecture that hit signal from multi-gap resistive plate chamber (MRPC) is signaled as LVDS by front-end electronics (FEE) and sent to the back-end time digitizing system via long shield differential twisted pair cables. The ETOF digitizing system consists of 2 VME crates each of which contains modules of time digitizing, clock, trigger and fast control etc. The time digitizing module (TDIG) of this prototype can support up to 72 electrical channels of hit information measurement. The fast control (FCTL) module can operate at barrel or endcap mode. The barrel FCTL fans fast control signals from the trigger system out to endcap FCTLs, merges data from endcaps and transfers to the trigger system. Without modifying the barrel TOF structure, this time digitizing architecture benefits for improving ETOF performance without degrading barrel TOF measuring. Lab experiments show that the time resolution of this digitizing system can be less than 20ps, and the data throughput to DAQ can be about 92Mbps. Beam experiments show that the complete time resolution can be less than 45ps.

## 1 Introduction[1]

Since the summer of 2008, BEPCII [1] started to work and the luminosity has been continuously improving to about $6.5\times10^{32} cm^{-2} s^{-1}$, which makes it possible for BESIII detector operating correctly and efficiently [2]. The BESIII TOF (time-of-flight) system is based on plastic scintillator bars read out by fast fine mesh photomultiplier tubes (PMT). It consists of a barrel and two endcaps [3]. The major physical goal of TOF is particle identification (PID) that plays an essential role for the study of τ-charm physics. The PID capability is determined by the resolution of time measurement. The current time resolution of TOF system is about 92ps in the barrel and 138ps in the endcaps for π, corresponding to the average K/π separation (2σ) around 1.1GeV [4]. To further improve the BESIII PID capability, the resolution of ETOF (endcap TOF) is planned to be upgraded to about 80ps, corresponding to 1.4GeV K/π separation (2σ), which will make BESIII be a state-of-the-art detector in the world and make it sufficient to identify charged particles over the entire momentum range of interest. The physical goal requires each ETOF channel having total time resolution better than 80ps among which only 25ps is caused by electronics.

To achieve this goal, in the upgrade program, the newly developed gaseous and widely used detector, multi-gap resistive plate chamber (MRPC) [5-7], is chosen for the ETOF detector, and an MRPC prototype with pad readout was developed. It is configured into a trapezium shape with 2×12 readout pads each of which supports signal reading out from both sides [8]. There will be 36 MRPCs for each endcap, corresponding to 864 electrical channels respectively. However, the original channel number before upgrade is only 48 for each endcap. The basic principle of upgrading is only to replace ETOF components with new ones while keeping other BESIII detector systems unchanged. Facing the pressure of front-end circuit size, design cost and complexity caused by sharply increased electrical channels, the lumped design methodology with long analogue signal (~18m) transmission adopted by barrel TOF should be given up and replaced with distributed one.

In the upgraded ETOF, the front-end electronics are separated from time digitizing system, moved to detector side and needed special characteristics to adapt the output weak charge signal from MRPC (about tens of *fC*). The ultra-fast and low power front-end amplifier-discriminator ASIC with 8 channels per chip and LVDS output driver, NINO [9], is the best choice for ETOF upgrading. It is designed for the use of MRPC detector. This distributed design methodology benefits for reducing time resolution degradation caused by electronics maximally. Combining with time over threshold (TOT) technique [10], the time (T) and charge (Q) information of hit signals captured by MRPC detector and digitized by HPTDC with 25ps bins [11] can be achieved together during one measurement. The leading edge position of the input signal represents the hit time and the width between leading and trailing edges represents the charge for PID. In the distributed scheme, signal from MRPC is amplified, shaped, discriminated, stretched and eventually converted into LVDS signal for long distance transmitting to back-end electronics for time measuring through high density digital cables. How to digitize signals from sharply increased front-end electrical channels becomes a critical design challenge for ETOF upgrade back-end readout electronics.

---

[1] * Supported by National Natural Science Foundation of China (10979003, 11005107)
1) E-mail: cping@ustc.edu.cn
2) E-mail: jixl@ihep.ac.cn

In this paper, a prototype of time digitizing system was proposed and developed for ETOF upgrade. To verify the performance of time measurement resolution and data transmission, some experiments were performed.

## 2 Architecture of Time Digitizing System

The original BESIII TOF readout system [12] consists of barrel and endcap TOF electronic modules that are housed in 2 VME crates, each of which contains 11 barrel and 3 endcap TOF-FEEs. The barrel and endcap modules are designed with the same circuit board formation. The operating mode can be switched by writing different values into an on-board register. Besides FEEs, there also contains FEE rear modules (FEE_Rear), fast control modules (FCTL) and clock modules. For the case of ETOF upgrading, signal from 1728 rather than 96 electrical channels need to be read out. ETOF readout modules have to be moved out from the original VME crates (now called BTOF crate) and rearranged into 2 new so-called ETOF crates, together with corresponding clock and fast control modules. After this rearrangement, the BTOF crate-A located on the 3$^{rd}$ floor of BESIII electronics building contains the original FEEs [12], FEE_Rears, TOF monitor [13], master clock [14] and PowerPC interface modules. The BTOF crate-B located on the ground floor keeps unchanged except for the lack of 3 ETOF modules.

As mentioned above, ETOF readout electronics has a distributed architecture. FEEs are moved to front-end detector side leaving time digitizing modules (called TDIG) residing in ETOF crates. For the purpose of interacting with trigger system, each ETOF crate also contains an extra fast control module besides TDIGs. For the purpose of system level clock synchronization, there also need a slave clock module in each ETOF crate to receive clock from the master clock module located in BTOF crate-A. To minimally modifying the original BESIII structure and to guarantee the system level reliability, the trigger system hardware and IO interfaces are unchanged except for minor logic and algorithm mending. The trigger system can only interact with 2 FCTL modules, so the FCTL module in BTOF crate-A should be redesigned and appointed as a router for transferring signals from FCTL modules in both ETOF crates to the trigger system, while the one in BTOF crate-B keeps unchanged.

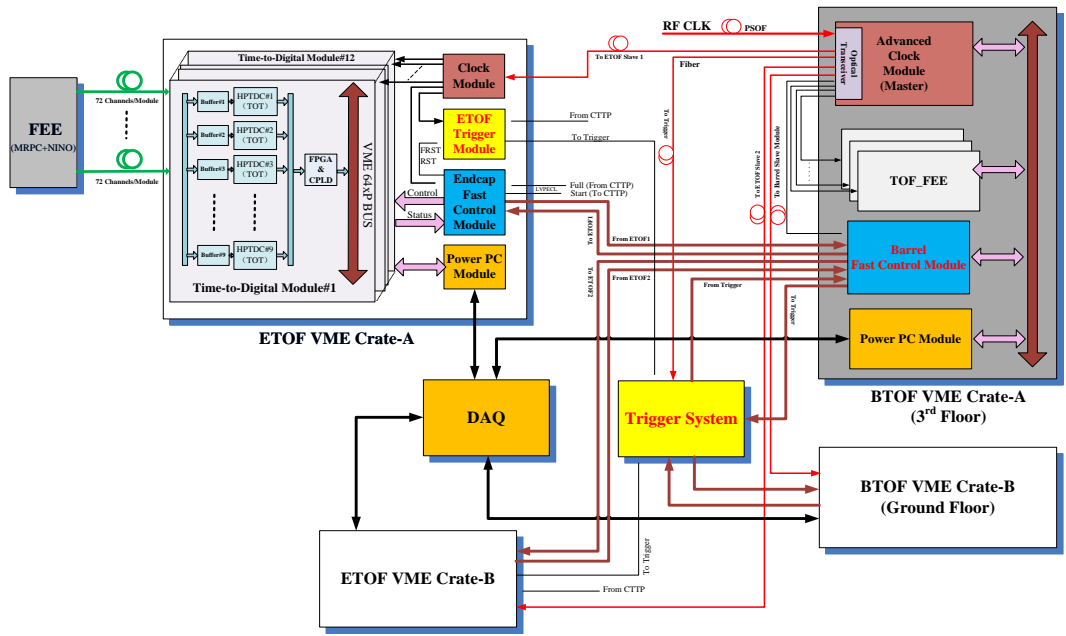

Fig. 1 The architecture of time digitizing system for ETOF upgrade

Fig. 1 schematically shows the structure and interfaces of ETOF time digitizing system planning to be upgraded. In this upgrading scheme, BTOF VME modules including FEE (rear), monitor, clock and fast control modules housed in crate-B placed on the ground floor can keep unchanged. The so-called advanced clock module [14] plugged into BTOF crate-A located on the $3^{rd}$ floor receives 500MHz RF clock from BEPCII accelerator via a phase stabilized optical fiber (PSOF) and executes tasks of TOF clock generating, synchronizing and distributing. It fans out 15 clocks signaled with LVPECL and 5 ones with optics. Slave clock modules plugged in ETOF crates receives the synchronized and stable clock from the master through optical fiber channel, and fans out 18 LVPECL clocks to feed modules inside the same ETOF crate and 2 optical clocks for backup or potential need of expansion.

Lacking connecting with FEEs, ETOF trigger selection module is removed from the original trigger system and implemented with VME module formation plugged into each ETOF crate. During operating, ETOF trigger module generates trigger condition according to signals from CTTP (coincide, threshold, test and power) modules and sends it to the general trigger system for trigger selection. Due to the way of MRPC installation differing from that of PMT, trigger algorithm corresponding to ETOF should be modified and implemented in ETOF trigger modules [15].

For the purpose of calibrating front-end electronics, ETOF FCTL module sends calibrating start signal to the front-end and then generates a pseudo L1 fast control signal to TDIG modules to simulate trigger arriving. Once receiving the start signal, each FEE will generate a step signal feeding to a discriminator to simulate a hit event. The delay between start and L1 leading edge can be adjusted by DAQ.

In this upgrade scheme, BTOF electronic modules almost keep unchanged, which guarantee the good performance of the original barrel TOF measurement and simplify the readout architecture and ETOF time digitizing system design meanwhile. To avoid the risk of clock performance degradation, the original BESIII clock distribution scheme is reserved with minor mending.

Compared with the original TOF readout electronics, there are 2 extra VME crates for ETOF upgrading. The front-end electronics and time digitizing system are separated and connected with each other via long cables. The sharply increased electrical channel number, detached architecture and newly designed VME modules make it necessary to evaluate some important system technical specifications, such as resolution of time measuring and ability of VME data transmitting with DAQ.

## 3 Time Digitizing Module Design

To evaluate the time digitizing resolution, a 9U VME prototype module called TDIG is developed [16]. Each TDIG module can receive 72 channels of time signal from 3 MRPC detectors via 3 individual high speed and high density shield differential twisted pair cables [17] respectively. The cable length is about 10m. Fig. 2 is the structure of a TDIG module.

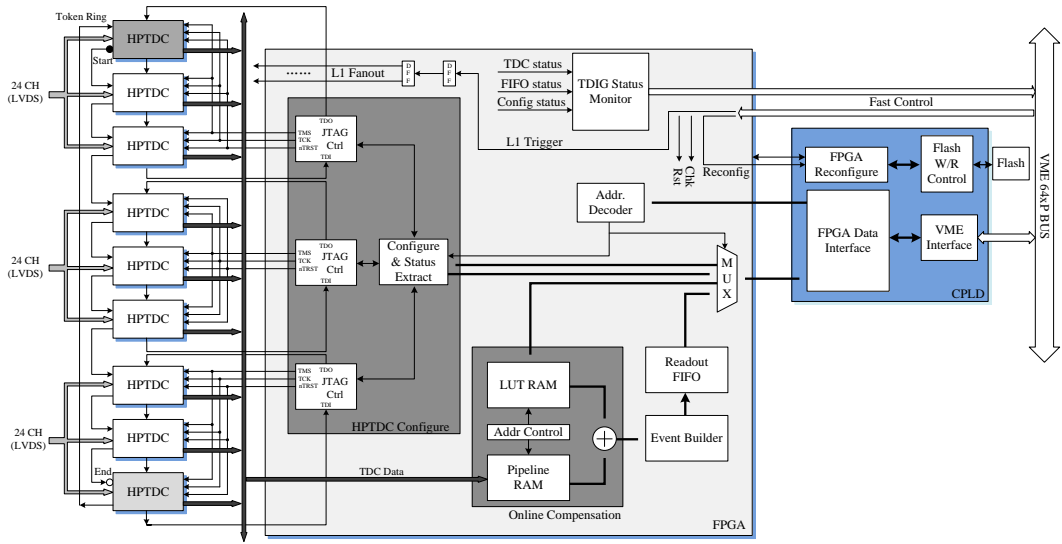

Fig. 2 Structure of TDIG module

To support up to 72 electrical channels, each TDIG needs 9 HPTDC totally. The TDCs are divided into 3 groups each of which is formed into a daisy chain respectively for the purpose of configuration or status loopback through JTAG interface. Configuring data sent from DAQ including setup or control registers is received via VME bus and buffered into a local memory in the Configure & Status Extract module. Once configuring command arrives, JTAG Ctrl sub-modules read out these data and feed to HPTDC via JTAG ports. These 3 JTAG Ctrl modules operate under the same rhythm, which makes these 3 groups of HPTDCs being configured simultaneously. In addition to configuring, JTAG Ctrl modules can also collect TDC status information including error, token, FIFO, trigger or DLL etc through JTAG ports. According to the need of experiment, status information, e.g. FIFO status indicating TDC overflow or not, can be fed to TDIG Status Monitor module to generate fast control signal sent to BESIII fast control system.

In the case of time information measurement, to avoid the LVDS signal quality degradation caused by long transmission cable and to improve the ability of driving TDCs, LVDS buffers (not showed in Fig. 2 for simplicity) are adopted before signal enters into TDC. After being configured correctly, HPTDC can enter into the very high resolution mode for time measuring with 25ps bin size. Once hit signal occurs on any HPTDC channel, it will be latched with coarse counter and fine calibrated values and stored into L1 buffer together with channel identifier waiting for L1 trigger selection. The L1 buffer, with depth of 256 words, is shared with 8 channels. The complete time measurement is encoded into a binary format. FCTL modules in BTOF or ETOF crates fans the L1 trigger signal generated from BESIII trigger system out to each TDIG module. Each TDIG receives L1 trigger signal from fast control system and fans further out to feed to the trigger input port of each HPTDC respectively. The mismatch between real physical event occurring (T0) and trigger input time due to trigger latency can be eliminated by a trigger matching mechanism that HPTDC searches hit information in L1 buffer and commits as valid only if this hit information is within a pre-set time range called match window [11].

In the case of data reading out, all HPTDCs are configured into a daisy-chain formation on which a token is transferred. For the purpose of fixing the token starting location, the HPTDC on the topside is appointed as the master TDC and numbered as 1. So the downside TDC is appointed as No. 9 where the token ends. All TDCs are configured into parallel readout mode that they share a command 32-bit bus for data reading out.

Benefitting from R-C delay line and data interpolation, HPTDC can achieve excellent measurement resolution of 25ps. But due to the process variations of the R-C delay line, such delay line has large dependencies on processing parameters and each TDC chip needs to be calibrated. The principle of calibration is to eliminate the INL in the raw data if the INL can be obtained prior and the value is fixed [16]. Actually, the R-C delay line in HPTDC has very small dependencies on temperature and supply voltage etc, which makes the INL of each channel be fixed. On the other hand, we can easily use software algorithm to compute and obtain the INL values corresponding to raw measurements without any calibration. In Fig. 2, we store the INL values obtained before calibrating into LUT-rams of FPGA through DAQ. During calibrating, TDIG will compensate each channel raw data online combining INL value corresponding to this channel.

The calibrated hit measurements are fed to event builder module for event building before being stored into event FIFO waiting for readout to DAQ via VME bus. To improve the reliability and portability, all operations corresponding to VME are implemented in a CPLD chip. Besides, for the purpose of adapting for flexible applications, CPLD also has ability of reconfiguring the FPGA logic. The new logic programming data is stored into a flash storage from DAQ.

## 4 Fast Control Module Design

Fast control (FCTL) system in TOF electronics fans fast control signals, such as L1 trigger, system reset sent from trigger system, out to each VME TOF module including BTOF FEE and ETOF TDIG. Meanwhile, FCTL collects status of TOF readout modules and sends back to global fast control system in trigger system. Fast control signals can be classified into 3 groups such that trigger, control and status signals. To support up to 4 individual VME crates containing all BTOF and ETOF readout modules, TOF FCTL system consists of 4 VME modules plugged into each crate respectively. Keeping BTOF readout modules on the ground floor of BESIII building unchanged, there are 3 newly developed FCTL modules such that 1 FCTL consider as master or router in BTOF crate-A ($3^{rd}$ floor) and 2 consider as slaves in 2 ETOF crates respectively. To simplify the design and use of FCTL module, master FCTL has the same hardware structure as that of slave one. Fig. 1 shows the connection between FCTL modules and the location of each module. The following Fig. 3 exhibits the structure of FCTL module.

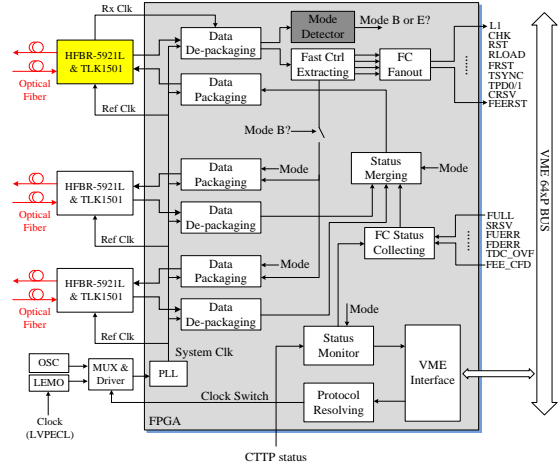

Fig. 3 Structure of fast control module

There are totally 4 optical fiber pairs with full duplex Rx and Tx channels. The FCTL can operate at two different mode, BTOF or ETOF mode. In the case of BTOF mode, 3 pairs are used and 1 pair is reserved for backup. While in the case of ETOF mode, 1 pair is used and 3 pairs are for backup. The topside optical pair is appointed as master pair for whichever mode. During operating, BTOF FCTL communicates with the trigger system through the master pair and ETOFs through 2 slave pairs. Unlike the original BTOF FCTL [12], the new FCTL module can switch to BTOF or ETOF mode automatically. Before operating, FCTL module detects mode, BTOF or ETOF, according to the mode information embedded into data stream it receiving. If it detects that the received data is sent from the trigger system, FCTL switches to BTOF mode. And if from BTOF FCTL, it switches to ETOF mode. In BTOF mode, FCTL extracts fast control information sent from trigger system and fans out to BTOF FEE modules via VME bus and to ETOF FCTL modules via the 2 slave optical fiber pairs embedded with BTOF mode information. In ETOF mode, FCTL extracts and fans out fast control to TDIG modules without fanning out via slave fiber pairs which are disabled instead. Whichever BTOF or ETOF FCTL, there is a module collecting fast control status sent from TOF readout modules. FC status from BTOF will be sent back to the trigger system merged with that of ETOFs. For the purpose of diagnosing or monitoring, FCTL or CTTP status monitoring data can be embedded into FC status stream or sent to DAQ from VME bus.

## 5 Experiments and Verification

The time measuring resolution plays essential role to the PID of BESIII TOF physical goal and is naturally appointed as a key TOF electronic specification for being improved continuously. Differing from BTOF, ETOF upgrading scheme adopts distributed design methodology while giving up the lumped one, which simplifies the FEE design but brings uncertainty about time measuring resolution meanwhile. The designed electronic resolution is about 25ps of RMS value. On the other hand, the sharply increased electrical channels raise the higher requirement for data transmission to DAQ than the case of BTOF. To evaluate the time digitizing resolution and data transmission performance, some experiments and verifications were carried out.

### 5.1 Data transmission evaluation

Due to the BESIII trigger latency about 6.4μs, to search the hit position precisely and efficiently, the match window of HPTDC is set to be 3μs that is nearly about half of the trigger latency. And due to the rare of multi-hitting in this time slot, the default hit number for each electrical channel is assigned as 1 while considering the output data amount. So we can obtain the following total output words (32bit) for each TDIG module:

72ch × hit occupancy × 2 × hit num + 4

where 2 denotes the leading and trailing edges for TOT application, 4 denotes the redundant information for data packaging. Table 1 shows the relationship between detector hit occupancy and ETOF electronic module data output.

Table 1 Relationship between occupancy and ETOF data

| hit occupancy (%) | hit ch nb per TDIG | Max event size per TDIG (byte) | Max event size per crate (byte) | Output data rate/crate (Mbytes/s) |
|---|---|---|---|---|
| 100 | 72 | 592 | 7104 | 28.416 |
| 50 | 36 | 304 | 3648 | 14.592 |
| 25 | 18 | 160 | 1920 | 7.68 |
| 11.1 | 8 | 80 | 960 | 3.84 |
| 5.6 | 4 | 48 | 576 | 2.304 |

For the case of 25% occupancy for ETOF, that is greater than the average level for BESIII experiment, the hit channel number of each TDIG module is 18 corresponding to event size 160 for each TDIG and 1920 for each ETOF crate. Due to the average trigger rate of 4000Hz, the data throughput for each crate should be greater than 7.68Mbyte/s. So a VME processor module with a 100Mbps Ethernet port, e.g. MVME5100, is enough for ETOF VME crate data readout. To read digitized hit information out as soon as possible, ETOF crate adopts chained block transfer (CBLT) DMA technique to transmit 12 TDIG data.

To evaluate the performance of VME data transmission, we designed the following test bench software shown in Fig. 4. There are 2 major modules such that command & network module and hardware interaction module in this software running on MVME5100 under the circumstance of VxWorks, a real-time embedded operating system. Cmd & Net module takes over the task of network communication and command executing, while the hardware interaction module, device driver related, takes over that of hardware (a VME bridge chip called universe2 that executes all VME bus operations) configuring, controlling and CBLT DMA handling.

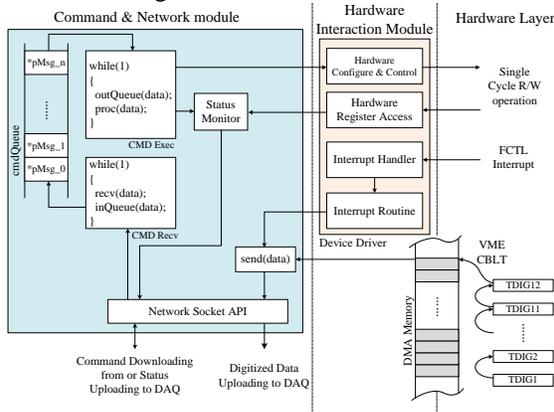

Fig. 4 Structure of VxWorks real-time software on VME single computer MVME5100

The experiment showed that the data throughput of VME bus can reach up to about 30Mbyte/s and that of 100M Ethernet interface on MVME5100 up to about 92Mbps. This result can meet the requirement of ETOF upgrade data transmission.

**5.2 Time digitizing INL evaluation**

To evaluate the time digitizing integral non-linearity (INL), we used a statistical code density test based on a source of random hits [18] and achieved the INL curve shown in Fig. 5. The INL is in the range of -2 ~9 LSB.

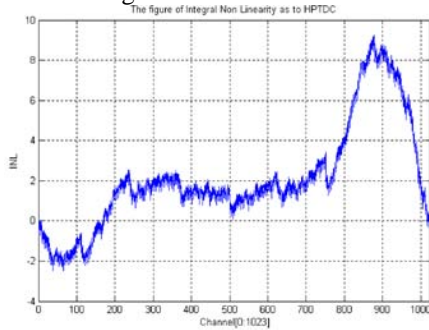

Fig. 5 INL curve of time digitizer

**5.3 Time digitizing resolution evaluation**

To evaluate time measurement resolution, we used the conventional cable delay method [16], with or without FEEs. Fig. 6 shows the time measurement resolution without FEEs while varying cable delay value for different HPTDCs. The test signal is generated from TEK AFG3252 and signaled to LVDS by a converter board. And the resolution is achieved by measuring the time interval for two different channels. According to Fig. 6, the resolution is better than 20ps that can meet the required 25ps for ETOF upgrade.

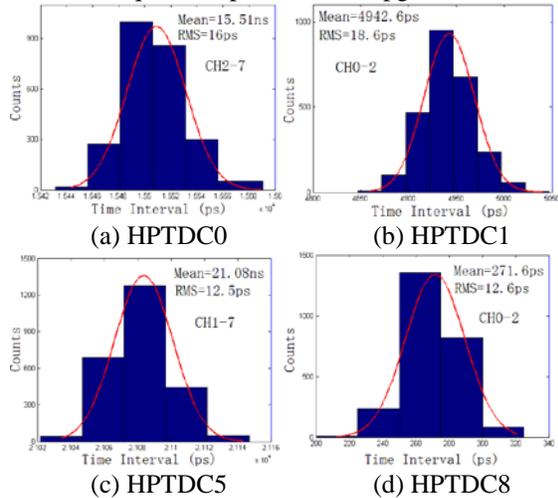

(a) HPTDC0   (b) HPTDC1

(c) HPTDC5   (d) HPTDC8

Fig. 6 Time measuring resolution for different HPTDCs (without FEE)

While evaluating measurement resolution with FEEs, we fed the test signal to FEEs for discriminating and LVDS signaling, measured the delay time of two randomly selected channels of one HPTDC and obtained the statistical histograms as shown in Fig. 7. Fig. 8 shows the obtained resolution of all HPTDC digitizing channels on one TDIG module.

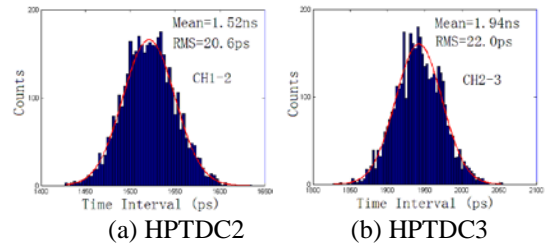

(a) HPTDC2   (b) HPTDC3

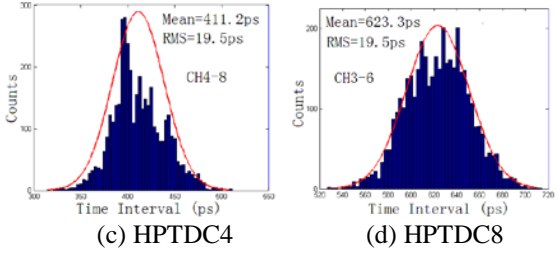

(c) HPTDC4   (d) HPTDC8

Fig. 7 Time measuring resolution for different HPTDCs (with FEEs)

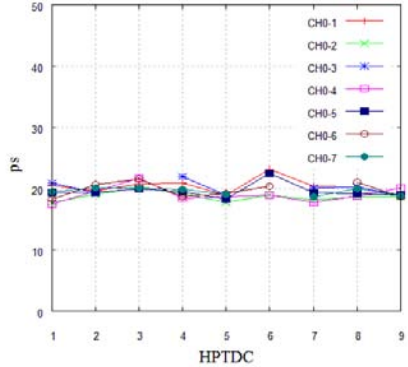

Fig. 8 Time measuring resolution (with FEEs) for all channels of a TDIG module

According to Fig. 7 and 8, we can conclude that this time digitizing prototype can achieve good measuring resolution better than 25ps with FEEs connected.

**5.4 Beam test**

Besides experiments above, we performed a beam test in IHEP, June 2011, to evaluate the complete time measurement resolution. The test configuration is shown in Fig. 9 schematically. To reduce the cosmic background, there located two scintillators along the beam direction. The coincidence will generate a signal used as the trigger for beam test if there is a valid beam signal passing through these two scintillators.

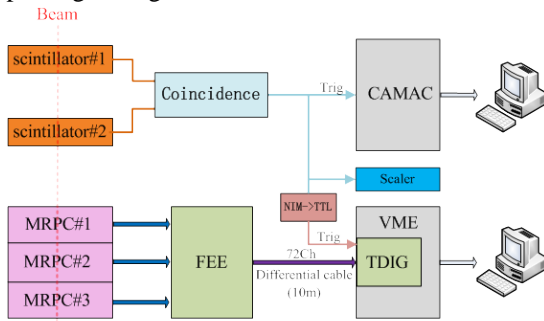

Fig. 9 Beam test configuration

Fig. 10 shows the complete measuring resolution [16], from which we can conclude that the time digitizing readout electronics can achieve good timing resolution 45ps that is better than 60ps that ETOF upgrade pursuing. Further, we can also conclude that the complete timing resolution is insensitive to the hit position.

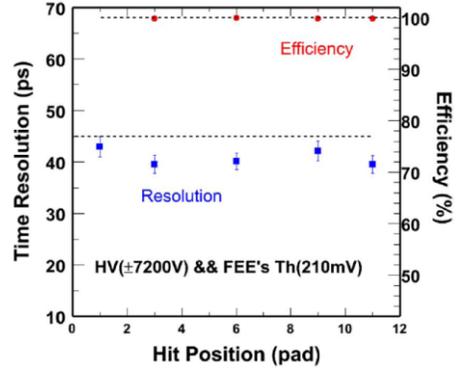

Fig. 10 Beam test result in June 2011

To evaluate the complete time measuring resolution contributed by electronics with or without beam, another beam test is performed in June 2012. The results are shown in Fig. 11 where marker (0) represents test without beam and (1) with beam. According to Fig. 11, we can conclude that the complete timing resolution contributed by electronics is better than 25ps except for that of HPTDC1. The measured resolution of even-to-even channels is obviously worse than that of even-to-odd channels. This is probably caused by improper electrical connections somewhere, because these are no any sighs exhibiting the failure of this HPTDC according to the abovementioned results. Next, some further beam tests will be performed to verify it.

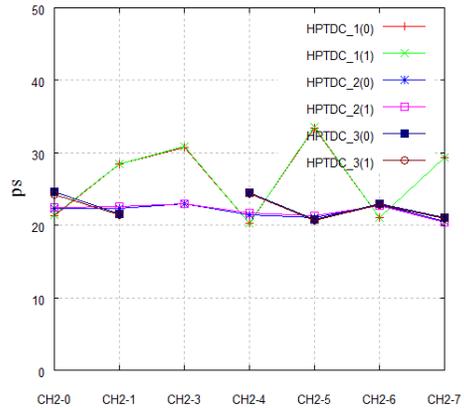

Fig. 11 Time measuring resolution contributed by electronics with or without beam

## 6 Conclusions

A prototype time digitizing system for BESIII ETOF upgrade is designed and introduced in this paper. The prototype has a formation of distributed structure which makes it possible and reliable to read signal out from sharply increased electrical channels of ETOF MRPC detectors via long differential twisted pair cables.

There exist two VME crates for ETOF upgrading, each of which contains 12 time digitizing modules (TDIG), 1 slave clock module, 1 fast control module (FCTL), 1 ETOF related trigger module and 1 bus controller module (MVME5100). All TDIG modules transfer digitized time information to the controller through CBLT DMA transaction initiated by FCTL. The meticulously designed embedded real-time VxWorks software on VME controller receives the digitized data from VME bus and transfers them to DAQ through 100M Ethernet.

To evaluate the performance of this prototype, experiments are performed with or without FEE, with or without beam. The experimental results show that the total data readout throughput of one VME crate can reach up to about 92Mbps that is similar to that of Daya bay DAQ [19-20]. The time measuring resolution can be better than 20ps without FEEs, 25ps with FEEs and 60ps with beam test. According to the experiments, we can conclude that this time digitizing prototype can meet the requirements for the physical goal of particle identification of ETOF upgrading.

Nevertheless, we will focus on the system integrating with the BTOF, trigger system and DAQ of BESIII, and focus on VME readout module engineering optimizing design including mechanical structure, reliability and consistency. Besides, we also need to manufacture modules and carry out short and long-term experiments and tests before assembling them into BESIII.

*We gratefully acknowledge Dr. Zhi Wu, Dr. Hongliang Dai and other members of BESIII ETOF upgrade group of IHEP for their earnest support and help during the beam tests.*